\documentclass[twocolumn,aps,superscriptaddress,showpacs]{revtex4}

\usepackage{amssymb}
\usepackage{amsmath}
\usepackage{mathrsfs}
\usepackage{graphicx}
\usepackage{subfigure}
\usepackage[normalem]{ulem}
\usepackage[dvips]{color}

\begin{document}

\title{Properties of quark matter and hybrid stars from a quasiparticle model}
\author{He Liu}
\email{liuhe@qut.edu.cn}
\affiliation{Science School, Qingdao University of Technology, Qingdao 266000, China}
\affiliation{The Research Center of Theoretical Physics, Qingdao University of Technology, Qingdao 266033, China}
\author{Yong-Hang Yang}
\affiliation{Science School, Qingdao University of Technology, Qingdao 266000, China}
\affiliation{The Research Center of Theoretical Physics, Qingdao University of Technology, Qingdao 266033, China}
\author{Yue Han}
\affiliation{Science School, Qingdao University of Technology, Qingdao 266000, China}
\affiliation{The Research Center of Theoretical Physics, Qingdao University of Technology, Qingdao 266033, China}
\author{Peng-Cheng Chu}
\email{kyois@126.com}
\affiliation{Science School, Qingdao University of Technology, Qingdao 266000, China}
\affiliation{The Research Center of Theoretical Physics, Qingdao University of Technology, Qingdao 266033, China}
\date{\today}
\begin{abstract}
 We investigate the properties of hybrid stars with the hadron-quark phase transition by using a quasiparticle model. Results from our study indicate that the coupling constant $g$ can stiffen the EOS of hybrid star matter and thus increase the hybrid star maximum mass and its tidal deformability, whereas it also decreases the mass and radius of the pure quark core. In addition, we find that a step change of the sound velocity occurs in the hadron-quark mixed phase, and it is restored with the decrease of nucleon and lepton degrees of freedom in the high density quark phase. The approximate rule that the polytropic index $\gamma \leq 1.75$  can also be used as a criterion for separating hadronic from quark matter in our work. The hypothesis of absolutely stable SQM (or "Witten hypothesis") suggests that a hybrid star containing a sufficient amount of SQM in its core will rapidly convert into a strange quark star. The SQM in hybrid stars therefore should break the absolutely stable condition, and the energy per nucleon ($E/A$) of both $ud$QM and SQM must exceed the lowest energy per nucleon 930 MeV. As a result, we provide the maximum mass, minimum radius $R_{1.4}$ and minimum tidal deformation $\Lambda_{1.4}$  of the hybrid stars as well as the maximum mass and radius of the quark matter core with different $g$ values within the allowable regions ($E/A>930$ MeV) on the $g-B^{1/4}$ plane. Using the constraints from astrophysical observations and heavy-ion experiments for comparison, our results indicate that the recently discovered massive neutron stars be well described as hybrid stars in the quasiparticle model, and confirm that the sizable quark-matter cores ($R_{QC}>6.5$ km) containing the mixed phase can appear in $2M_{\odot}$ massive stars.
\end{abstract}


\maketitle
\section{Introduction}
\label{INT}
Understanding the nature of the strongly interacting matter, especially the phase transition, is a major thrust of current research in both nuclear physics and astrophysics. It is generally believed that a first-order phase transition from the hadronic to quark matter at high baryon densities may occur in the interior of massive neutron stars (NSs)~\cite{Gle01,Web05,Aok06}. Some recent studies, for instance, have also shown that quark-matter cores can appear in massive neutron stars~\cite{Ann20}, and the presence of a first-order phase transition from hadronic to quark matter can imprint signatures in binary mergers observations~\cite{Alf19,Bau19}, as well as the sudden decrease in the gravitational-wave frequency in the binary-neutron-star (BNS) merger simulations is closely related to the hadron-quark phase transition~\cite{Hua22}. The appearance of quark matter in massive stars is considered as a hot topic in compact object studies, and the neutron stars with a hadron mantle and a quark core is usually referred to as hybrid stars. Hence, observations of massive neutron stars can provide us with a valuable window into an otherwise inaccessible realm of quark matter.

 Some inspiring progress over the last several years have been made in neutron star observations, and thus in our understanding of the equation of state (EOS) of neutron star matter. The measurements of PSR J1614-2230 and PSR J0348+0432 had led to a precise determination of $1.908\pm0.016M_{\odot}$~\cite{Dem10,Fon16,Arz18} and $2.01\pm0.04M_{\odot}$~\cite{Ant13} for their respective masses. More recently, the first simultaneous measurements of the mass and radius of a neutron star using the Neutron Star Interior Composition Explorer (NICER) data were those of the millisecond pulsar PSR J0030+0451. The two independent analyses predict ($68\%$ credible interval) $M=1.34^{+0.15} _{-0.16} M_{\odot}, R=12.71^{+1.14}_{-1.19}$ km~\cite{Ril19} and $M=1.44^{+0.15}_{-0.14} M_{\odot}, R=13.02^{+1.24} _{-1.06}$ km~\cite{Mil19}. PSR J0740+6620 has a gravitational mass of $2.08\pm0.07M_{\odot}$, which is considered as the highest reliably determined neutron star mass~\cite{Cro20,Ril21,Mil21}. Its radius were determined using the NICER and X-ray Multi-Mirror (XMMNewton) data with the results for the radius $12.39^{+1.30}_{-0.98}$ ~\cite{Ril21} and $13.71^{+2.61}_{-1.50}$ km~\cite{Mil21}  ($68\%$  credible interval). And the radius range that spans the $\pm1\sigma$ credible intervals of all the radius estimates in the different frameworks is $12.45\pm 0.65$ km for a canonical mass $M=1.4M_{\odot}$ neutron star~\cite{Mil21}. The gravitational wave events GW170817~\cite{Abb17} and GW190814~\cite{Abb20} have provided more additional constraints on the EOS of neutron star matter. The analysis of GW170817 by the LIGO/Virgo Collaboration has found with a $90\%$ confidence that the tidal deformability of the merging neutron stars constrained as the range $70 <\Lambda_{1.4}<580$~\cite{Abb18}.  Besides, the newly discovered neutron binary merger GW190814 which has a secondary component of mass $(2.50 \sim 2.67) M_{\odot}$ at $90\%$ credible level has also aroused lots of debates on whether the candidate for the secondary component is a neutron star or a light black hole~\cite{Abb20}. These observation events/objects comprise the multi-messenger data set for our following analyses on the properties of neutron star matter.

 The existence of such high-mass neutron stars indicates that the EOS of neutron star matter is relatively stiff, whereas the tidal deformation in gravitational wave observations of GW170817 and GW190425 implies a soft EOS at the intermediate density range~\cite{Mil21}. The measurement of the radius of a $1.4M_{\odot}$  star has its greatest impact on our understanding of matter at 1.6 times saturation density ($\rho_0 \approx 0.16$ fm$^{-3}$),  and the measurement of the radius of a $2.0M_{\odot}$  star can inform us primarily about matter at density $2.2\rho_0$~\cite{Dri21,Xie20}. Besides, the EOS of strongly interacting matter at densities $2\rho_0 <\rho <5\rho_0$ has also been constrained by the measurements of collective flows~\cite{Dan02} and subthreshold kaon production~\cite{Fuc06} in relativistic heavy-ion collisions. When taken together, these observations and experimental information have already narrowed significantly the EOS range of allowed theoretical models. Recent work on the EOS by effective models has generally concluded that the EOS of neutron star matter must be moderately soft at intermediate densities and stiff enough at high densities, which is highly relevant to a possible phase transition to quark matter~\cite{Alf13,Mon19,Lij20,Znb21,Lij22,Liu22}.

 In the present study, the neutron stars could be converted to hybrid stars with the hadron-quark phase transition. We describe strange quark matter (SQM) in hybrid stars based on the quark quasiparticle model, and nuclear matter using an improved isospin- and momentum-dependent interaction (ImMDI) model. The ImMDI model is constructed from fitting cold nuclear matter properties at saturation density and the empirical nucleon optical potential~\cite{Jxu15,Jxu19}, and it has been extensively used in intermediate energy heavy-ion reactions to study the properties of nuclear matter. The Gibbs construction~\cite{Gle92} is adopted for the description of hadron-quark mixed phase, where the coexisting hadronic and quark phases need to satisfy the $\beta$-equilibrium and charge-neutral conditions. While previous studies have explored the equation of state (EOS) of hybrid star matter using various models with the Gibbs construction and modified bag constants~\cite{Fis10,Jxu10,Wei11}, our focus is on examining the global properties of hybrid stars and quark-matter cores by excluding absolutely stable SQM inside hybrid star. This article is organized as follows. In Section II, we describe the quasiparticle model with different parameter sets for the quark matter at zero temperature. We present in Section III that the absolute stabilities of SQM and $u-d$ quark matter ($ud$QM); the properties of hybrid star matter, such as the EOS, the sound velocity $c_s^2$, and the polytropic index $\gamma$; the mass-radius relation and the dimensionless tidal deformability of hybrid stars, as well as the  mass-radius relation of quark-matter cores.  Our conclusions are given in Section IV.

\section{The theoretical model }
\label{MODEL}
The hybrid EOS consists of a hadronic phase connected to a quark phase through a hadron-quark mixed phase. The possible appearance of hyperons is neglected, which is due to the fact that there are still large uncertainties on the hyperon-nucleon ($YN$) and hyperon-hyperon ($YY$) interactions in the nuclear medium~\cite{Hiy20,Con16}. Besides, following the results from Ref.~\cite{Jxu10}, the fraction of hyperons disappears quickly in hadron-quark mixed phase, which means that the effect of hyperons at high densities, especially in the hadron-quark mixed phase, is expected to be small. Thus, we mainly focus on the properties of hybrid star matter without hyperons in this work.

Here we want to apply an ideal gas of quasiparticles with effective masses to the case of strange quark matter. The effective quark masses are derived from the zero momentum limit of the dispersion relations following from an effective quark propagator obtained from resumming one-loop self energy diagrams in the hard dense loop (HDL) approximation at finite chemical potential~\cite{Man96}. And the effective quark mass for each flavor of quarks in quasiparticle model at zero temperature can be expressed as~\cite{Sch97,Wen09,Chu21}
\begin{eqnarray}
m_q=\frac{m_{q0}}{2}+\sqrt{\frac{m_{q0}^2}{4}+\frac{g^2\mu_q^2}{6\pi^2}},
\end{eqnarray}
where $m_{q0}$ is the current mass for three flavor quarks, which is set as $m_{u0} = 5.5$ MeV, $m_{d0} = 5.5$ MeV, and $m_{s0} = 95$ MeV respectively in this work~\cite{Chu21,Zha21}. $\mu_q$ means the chemical potential for the different flavor of quarks, and $g$ represents the coupling constant of the strong interaction. In principle the value of $g$ should be determined as a $\mu$-dependent running coupling constant from the renormalisation group equation at finite density. However, so far there are no clear results at finite temperature or density~\cite{Eij94}, and thus $g$ is taken as a free parameter ranging from 1 to 5.

The quasiparticle contribution to the thermodynamic potential density for SQM can be written as
\begin{eqnarray}
\Omega=\sum_{q=u,d,s}[\Omega_q+B_q(\mu_q)]+B,
\end{eqnarray}
where $\Omega_q$ in the sum shows the contribution to the thermodynamic potential density for all flavors of quarks ($u$, $d$, and $s$), $B_q(\mu_q)$ is the additional terms for quarks, which is defined as a necessary energy counterterm in order to maintain thermodynamic self-consistency~\cite{Gor95}, and $B$ denotes the phenomenological bag constant which corresponds to the negative vacuum pressure term for nonperturbative confinement~\cite{Pat96}. The expression of $\Omega_q$ at temperature $T=0$ and chemical potential $\mu_q$ for the quasiparticle of mass $m_q$ can be written as
\begin{eqnarray}
\Omega_q=-\frac{d_q}{24\pi^2}[\mu_qk_{Fq}(\mu_q^2-\frac{5}{2}m_q^2)+\frac{3}{2}m_q^4\textrm{ln}(\frac{k_{Fq}+\mu_q}{m_q})],
\end{eqnarray}
where $d_q$ denotes the degree of degeneracy (e.g. $d_q= 6$ for quarks), and the Fermi momentum is $k_{Fq} = (\mu_q^2 -m_q^2)^{1/2}$. The term $B_q (\mu_q)$ is determined as
\begin{eqnarray}
B_q (\mu_q)&=&-\int\frac{\partial\Omega_q}{\partial m_q}\frac{\partial m_q}{\partial \mu_q}d\mu_q.
\end{eqnarray}

The the pressure $P_Q$ and energy density $\varepsilon_Q$ of SQM are, respectively, given by
\begin{eqnarray}
P_Q&=&-\sum_{q=u,d,s}[\Omega_q+B_q(\mu_q)]-B,
\\
\varepsilon_Q&=&\sum_{q=u, d, s}\mu_q\rho_q-P_Q,
\end{eqnarray}
where $\rho_q$ stands for the net quark number density. In the quark phase, the system is composed of a mixture of quarks ($u$, $d$, and $s$) and leptons ($e$ and $\mu$) under the charge neutrality condition
\begin{equation}
\frac{2}{3}\rho_u-\frac{1}{3}(\rho_d+\rho_s)-\rho_e-\rho_\mu=0,
\end{equation}
and the $\beta$-equilibrium condition
\begin{eqnarray}
\mu_s&=&\mu_d=\mu_u+\mu_e,
\\
\mu_\mu&=&\mu_e.
\end{eqnarray}
In terms of the electron mass $m_e=0.511$MeV and the muon mass $m_\mu=106$ MeV, the lepton contributions to the energy density and the pressure are
\begin{eqnarray}
\varepsilon_L&=&\sum_{i=e, \mu}\frac{1}{\pi^2}\int_0^{k_{Fi}}\sqrt{p^2+m_i^2}p^2dp,
\\
P_L&=&\sum_{i=e, \mu}\mu_i\rho_i-\varepsilon_L,
\end{eqnarray}
where $k_{Fi}=(3\pi^2\rho_i)^{\frac{1}{3}}$ is the lepton fermi momentum. The total energy density and pressure including the contributions from both quarks and leptons in quark phase are given by
\begin{eqnarray}
\varepsilon^Q&=&\varepsilon_Q+\varepsilon_L,
\\
P^Q&=&P_Q+P_L.
\end{eqnarray}

In the hadronic phase, an improved isospin- and momentum-dependent interaction (ImMDI) model is used to describe the $\beta$-equilibrium and charge-neutral nuclear matter. In our previous study~\cite{Liu22}, the ImMDI model is fitted to the properties of cold symmetric nuclear matter (SNM), which is approximately reproduced by the self-consistent Greens function (SCGF) approach~\cite{Car14,Car18}  or chiral effective many-body pertubation theory ($\chi$EMBPT)~\cite{Wel15,Wel16}. And the parameters $x$, $y$ and $z$ are introduced to adjust the slope $L$ of symmetry energy, the momentum dependence of the symmetry potential, and the symmetry energy $E_{sym}(\rho_0)$ at saturation density, respectively. Recently, the discovery of GW170817 has triggered many analyses of neutron star observables to constrain nuclear symmetry energy. The average value of the slope parameter of the symmetry energy $L$ from the 24 new analyses of neutron star observables since GW170817 was about $L= 57.7 \pm 19$ MeV at a $68\%$ confidence level~\cite{BLi21}, which is consistent with the latest report of the slope parameter $L$ between 42 and 117 MeV from studying the pion spectrum ratio in heavy-ion collision in an experiment performed at RIKEN~\cite{Est21}. However, the Lead Radius Experiment (PREX-II) reported very recently new constraints on the neutron radius of $^{208}$Pb, which implies a neutron skin thickness of $R^{^{208}Pb}_{skin} = 0.283 \pm 0.071$ fm~\cite{Adh21} and constrains the slope parameter to $L= 106 \pm 37$ MeV~\cite{Ree21}, which is much larger than many previous constraints from microscopic calculations or experimental measurements~\cite{Tsa12,Lat13,BLi21}.  In order to better focus on the properties of quark matter, we thus choose for the hadronic phase a fixed parameter set, $x=-0.3$, $y=32$ MeV, and $z=0$, that would allow $2.08M_{\odot}$ neutron stars and still satisfy well nuclear matter constraints at saturation density, i.e., the binding energy $E_0(\rho_0) = -15.9$ MeV, the incompressibility $K_0 = 240$ MeV, the symmetry energy $E_{sym}(\rho_0) = 32.5$ MeV, the slope parameter $L=106$ MeV, the isoscalar effective mass $m^{\star}_s = 0.7m$, and the single-particle potential $U_{0,\infty} = 75$ MeV at infinitely large nucleon momentum.

The hadron-quark mixed phase is predicted to exist in the region between hadronic matter and quark matter based on various theoretical approaches. In the Maxwell construction, the coexisting hadronic and quark phases have equal pressure and baryon chemical potential but different electron chemical potential. The Gibbs construction is more generally adopted for the description of hadron-quark mixed phase, where the coexisting hadronic and quark phases are allowed to be charged separately. Besides, the mixed phase in the Gibbs construction persists within a limited pressure range, so it is convenient to form a massive neutron star containing the mixed phase. Both of the Maxwell and Gibbs constructions involve only bulk contributions, but the finite-size effects like surface and Coulomb contributions are neglected. The possible geometrical structure of the mixed phase has been extensively discussed in Refs.~\cite{Mar07,Na12,Wu19,Lug21,Ju21}. However, the large uncertainties in the structure and density range of the mixed phase are still present. In the present work, the hadron-quark mixed phase is described by imposing the Gibbs construction~\cite{Gle92,Gle01}: $T^H=T^Q$, $P^H=P^Q$, $\mu_B^H=\mu_B^Q$, and $\mu_c^H=\mu_c^Q$, where $\mu_B$ and $\mu_c$ are the baryon and charge chemical potential, as well as the labels $H$ and $Q$ represent the hadronic and quark phases, respectively. Adding baryon number conservation and charge neutrality conditions, the dense matter enters the mixed phase, in which the hadronic and quark matter need to satisfy following equilibrium conditions:
\begin{eqnarray}
\mu_i&=&\mu_Bb_i-\mu_cq_i,  \quad P^H=P^Q,
\notag\\
\rho_B&=&(1-Y)(\rho_n+\rho_p)+\frac{Y}{3}(\rho_u+\rho_d+\rho_s),
\notag\\
0&=&(1-Y)\rho_p+\frac{Y}{3}(2\rho_u-\rho_d-\rho_s)-\rho_e-\rho_\mu,
\end{eqnarray}
where $Y$ is the baryon number fraction of the quark phase. The crust of hybrid stars, in our calculations, is considered to be divided into two parts: the inner and the outer crust as in the previous treatment~\cite{Jxu09C, Jxu09J}. The polytropic form $P = a + b\varepsilon^{4/3}$ has been found to be a good approximation to the inner crust EOS~\cite{Car03}, and the outer crust usually consists of heavy nuclei and electron gas, where we use the EOS in Ref.~\cite{Bay71}.

Using the whole EOS from hadronic to quark phase, the mass-radius relation of hybrid stars can be obtained by solving the Tolman-Oppenheimer-Volkoff equation, which can be written as
\begin{eqnarray}
\frac{dP(r)}{dr}&=&-\frac{M(r)[\varepsilon(r)+P(r)]}{r^2}[1+\frac{4 \pi P(r)r^3}{M(r)}]
\notag\\
&\times&[1-\frac{2M(r)}{r}]^{-1},
\end{eqnarray}
where $\varepsilon(r)$ is the energy density and $P(r)$ is the pressure obtained from the equation of state. $M(r)$ is the gravitational mass inside the radius $r$ of the hybrid star given by
\begin{eqnarray}
\frac{dM(r)}{dr}&=&4 \pi r^2 \varepsilon(r).
\end{eqnarray}

The gravitational waves emitted from the merge of two neutron stars are considered as another probe to the EOS of dense matter~\cite{Hin08,Rea09}. The tidal deformability $\Lambda$ of neutron stars during their merger is related to the Love number $k_2$ through the relation $k_2 =3/2\Lambda\beta^5$~\cite{Hin08,Pos10}, which can be given by
\begin{eqnarray}
k_2 &=&\frac{8}{5}\beta^5(1-2\beta)^2[2-y_R+2\beta(y_R-1)]
\notag\\
&\times& \{2\beta[6-3y_R+3\beta(5y_R-8)]
\notag\\
&+& 4\beta^2[13-11y_R+\beta(3y_R-2)+2\beta^2(1+y_R)]
\notag\\
&+&3(1-2\beta)^2[2-y_R+2\beta(y_R-1)]\text{ln}(1-2\beta)\}^{-1},
\end{eqnarray}
where $\beta \equiv M/R $ is the compactness of the star, and $y_R \equiv y(R)$ is the solution at the neutron star surface to the first
order differential equation
\begin{eqnarray}
r\frac{dy(r)}{dr}+y(r)^2+y(r)F(r)+r^2Q(r)=0,
\end{eqnarray}
with
\begin{eqnarray}
F(r) &=& \frac{r-4\pi r^3[\varepsilon(r)-P(r)]}{r-2M(r)},
\notag\\
Q(r)&=&\frac{4\pi r[5\varepsilon(r)+9P(r)+\frac{\varepsilon(r)+P(r)}{\partial P(r)/\partial \varepsilon(r)}-\frac{6}{4\pi r^2}]}{r-2M(r)}
\notag\\
&-&4[\frac{M(r)+4\pi r^3P(r)}{r^2(1-2M(r)/r)}]^2.
\end{eqnarray}
For a given central density $\rho_c$ and using the boundary conditions in terms of $y(0) = 2$, $P(0)=P_c$, $M(0)=0$ and $\varepsilon(0)=0$, the mass $M$, radius $R$, and the tidal deformability $\Lambda$ can be obtained once an EOS is supplied.

\section{Results and Discussions}
\label{RAD}
The absolutely stable condition of SQM/$ud$QM has been proposed in Ref.~\cite{Far84}, which can put very strict constraints on the parameter space for most of the phenomenological quark matter models. The ordinary nuclei are made of nucleons and not of a two-flavor quark phase, the energy per nucleon ($E/A$) of $ud$QM therefore must exceed the lowest energy per nucleon found in nuclei, which is about 930 MeV for iron ($M(^{56}Fe/56)$). For the strange quark matter, Ref.~\cite{Far84} points out SQM in aggregates large enough that surface effects can be ignored and that electrons (or positrons) bound to it by Coulomb forces are inside the chunk and numerous enough to be treated as a degenerate Fermi gas. This requires $E/A$ of SQM to be less than that of the nucleon ($M_N=939$ MeV). Actually for $E/A$ between 930 and 939 MeV SQM could decay by emission of nuclei accompanied by weak interactions to maintain flavor equilibrium, which means that the minimum value of $E/A$ of the absolutely stable SQM at zero temperature should be less than 930 MeV. However, the hypothesis of absolutely stable SQM (or "Witten hypothesis") suggests that the energy required to create a single strange quark could be offset by the energy released when two up or down quarks combine with a strange quark to form a lower-energy state, which would imply that the absolutely stable SQM is the ground state of matter, and that all forms of matter would eventually decay into SQM~\cite{Wit84,Bay85,Oli87}. In this case, a hybrid star containing a sufficient amount of SQM in its core will rapidly convert into a strange quark star. Thus, the SQM in hybrid stars should break the absolutely stable condition, and the energy per nucleon ($E/A$) of both $ud$QM and SQM must exceed the lowest energy per nucleon 930 MeV.

\begin{figure}[tbh]
\includegraphics[scale=0.30]{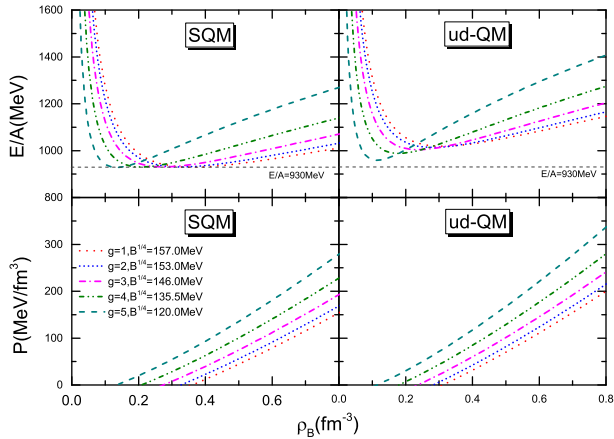}
\caption{(color online) The energy per nucleon (E/A) and the corresponding pressure as functions of the baryon density for SQM (left panels) and $ud$QM (right panels) at zero temperature from the quasiparticle model with different parameter sets. The horizontal dashed line $E/A$=930 MeV is also shown for comparison. } \label{fig1}
\end{figure}

We first present in Fig.~\ref{fig1} the energy per nucleon and corresponding pressure of SQM and $ud$QM as functions of baryon number density in the quasiparticle model with five different parameter sets: $g=1, B^{1/4}=157.0$ MeV; $g=2, B^{1/4}=153.0$ MeV; $g=3, B^{1/4}=146.0$ MeV; $g=4, B^{1/4}=135.5$ MeV; and $g=5, B^{1/4}=120.0$ MeV. These parameter sets include various values for the coupling constant $g$ and bag constant $B^{1/4}$. However, it should be noted that all the minimum values of energy per nucleon of SQM and $ud$QM are larger than 930 MeV in all of the five parameter settings, which is due to the break in absolute stability for quark matter. We have ensured that the minimum value of energy per nucleon of SQM is only slightly larger than 930 MeV in our parameter settings, but the minimum value of energy per nucleon actually increases with increasing bag constant $B$ while fixing the coupling constant $g$. Therefore, the value of $B$ in our present parameter sets is the minimum value. It is important to consider that the ranges of $B$ and $g$ can significantly impact the properties of SQM and hybrid stars, which will be further discussed later in this work. Furthermore, it can be seen from Fig.~\ref{fig1} that the baryon density of the minimum energy per nucleon for SQM and $ud$QM in all cases corresponds exactly to the zero pressure points, satisfying the thermodynamic self-consistency of quark matter. Although the baryon density of the zero pressure point decreases with the coupling constant $g$, the equation of state containing SQM/$ud$QM becomes stiffer, which can support more massive hybrid stars.

\begin{figure}[tbh]
\includegraphics[scale=0.30]{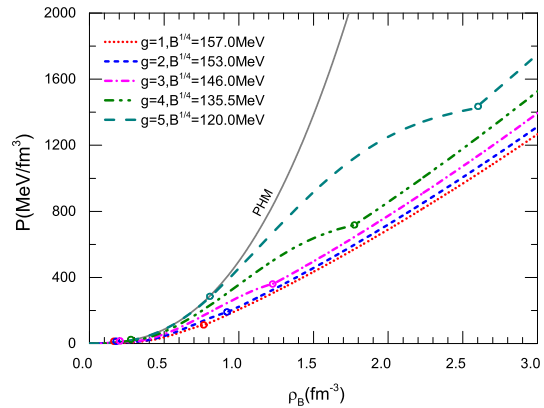}
\centering
\caption{(color online) The equation of state (EOS), pressure as a function of the baryon density, of hybrid star matter based on the ImMDI interaction for nuclear matter with a fixed parameter set ($x=-0.3$, $y=32$ MeV, and $z=0$) and the qusiparticle model for SQM with different parameter sets. The two cycles with same color represent the the range of the hadron-quark mixed phase, and the result from pure hadronic matter (PHM) is also shown for comparison.}\label{fig2}
\end{figure}

We show in Fig.~\ref{fig2} the EOS of hybrid star matter with the hadron-quark phase transition in different parameter sets. The ImMDI interaction with a fixed parameter set, $x=-0.3$, $y=32$ MeV, and $z=0$, is used for nuclear matter, and the two cycles with same color in Fig.~\ref{fig2} represent the the range of the hadron-quark mixed phase. It can be seen that the phase transition leads to a softening of EOS of hybrid star matter, compared to their purely hadronic counterpart. Moreover, the EOS of hybrid star matter is sensitive to the strength of the coupling constant $g$. With increasing the coupling constant $g$ for SQM the EOS of hybrid star matter becomes stiffer, which is consistent with the results of strange quark star (QS) matter in Ref.~\cite{Zha21}, and the onset of the phase transition is moving to higher densities since the transition pressure is also increasing under the Gibbs construction. It should be noted that the phase transition in most cases occurs at density $\rho_0 \sim 5\rho_0$, where the EOS of SQM are considered as an important factor affecting the properties of hybrid stars~\cite{Xie20,Liu22}. On the other hand, the stiffened EOS can increase the baryon density of the hadron-quark phase transition, which lead to the gradual weakening of the role of quark matter for hybrid stars.

Compared to the hadronic matter (HM), SQM is known to exhibit markedly different properties. For example, SQM at very high densities ($\rho_B\geq 40\rho_0$) is approximately scale-invariant or conformal, whereas in HM the degree of freedom is smaller and the scale invariance is also violated by the breaking of chiral symmetry. These qualitative differences between HM and SQM can be reflected in the different physical quantities. The sound velocity $c_s$, which can be calculated from $c_s^2 = \partial{P}/\partial{\varepsilon}$,  takes the constant $c_s^2 =1/3$ in the exactly conformal matter corresponding to SQM at high densities. However, $c_s^2$ in HM varies considerably: below saturation density, most hadronic models, such as chiral effective field theory, indicate $c_s^2\ll1/3$, while at higher densities the maximum of $c_s^2$ is predicted to be greater than 0.5~\cite{Gan09,Tew13}. Another physical quantity is the polytropic index $\gamma=d(\textrm{ln}P)/d(\textrm{ln}\varepsilon)$,  which is considered to be a good approximate criterion for the evidence of SQM in neutron stars.  The polytropic index has the value $\gamma = 1$ in conformal matter, while the hadronic models generically predict $\gamma\approx 2.5$ around saturation density~\cite{Kur10}.

\begin{figure}[tbh]
\includegraphics[scale=0.26]{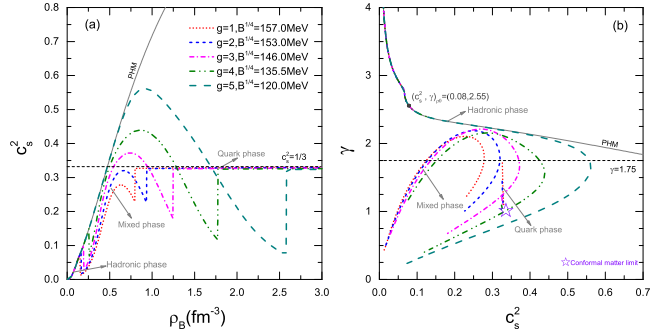}
\centering
\caption{(color online) The squared speed of sound $c_s^2$  as a function of the baryon density (panel (a)), and the relation between the polytropic index $\gamma$  and the squared speed of sound $c_s^2$  (panel (b)) for hybrid star matter by varying the coupling constants $g$ from the quasiparticle model. The dash lines $c_s^2=1/3$ and $\gamma=1.75$, as well as the violet star indicated the high-density conformal matter limit are also shown for comparison.}\label{fig3}
\end{figure}

 In Fig.~\ref{fig3} (a), we show that the squared speed of sound $c_s^2$  as a function of the baryon density for hybrid star matter with the hadron-quark phase transition by varying the coupling constants $g$ from the quasiparticle model. With the increase of $g$ in the hadron-quark mixed phase, the curve increases rapidly and reaches a larger peak. Meanwhile, we can see that $c_s^2$ in quark phase is insensitive to $g$, and slowly approaches the value $c_s^2 = 1/3$ of conformal matter in the high density. It should be noted that a step change of the sound velocity occurs in the hadron-quark phase transition where the quarks appear and thus soften the EOS as a result of more degrees of freedom, and it is restored with the decrease of nucleon and lepton degrees of freedom in the high density quark phase. Further, the step change of the sound velocity in hadron-quark phase transition is relevant to the frequency of the main peak of the postmerger gravitational wave (GW) spectrum ($f_2$), which is expected to be confirmed by future kilohertz GW observations with third-generation GW detectors~\cite{Hua22}. In Fig.~\ref{fig3} (b), we show the relation between the polytropic index $\gamma$ and the squared speed of sound $c_s^2$ in hybrid star matter. One can be seen that at saturation density $c_s^2$ and $\gamma$ are respectively 0.08 and 2.55 from the ImMDI model with a fixed parameter set, $x=-0.3$, $y=32$ MeV, and $z=0$, which are consistent with those in most hadronic models. Both $c_s^2$ and $\gamma$ in the high density quark phase approach the conformal matter limit. Our results also agree with the approximate rule following Ref.~\cite{Ann20} that the polytropic index $\gamma \leq 1.75$ can be used as a criterion for separating hadronic from quark matter.

\begin{figure}[tbh]
\includegraphics[scale=0.30]{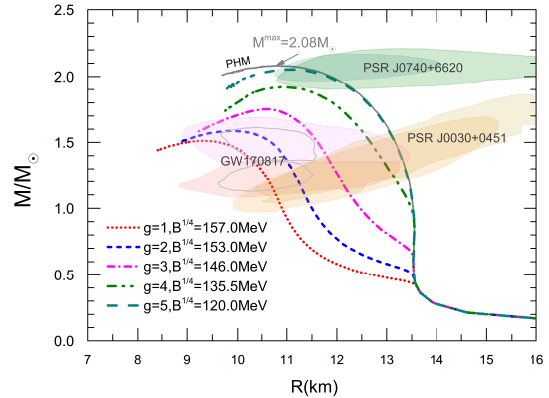}
\caption{(color online) Mass-radius relation of hybrid stars based on the quasiparticle model for SQM with the different coupling constants $g$. Constraints from multi-messenger astronomy observations~\cite{Abb17,Ril19,Ril21,Mil19,Mil21} are shown by shaded regions, see text for details.} \label{fig4}
\end{figure}

The mass-radius relation of hybrid stars is shown in Fig.~\ref{fig4}, based on the quasiparticle model for SQM with the different coupling constants $g$. The constraints from the bayesian analyses of the observational data from the pulsars PSR J0030+0451~\cite{Ril19,Mil19} and PSR J0740+6620~\cite{Ril21,Mil21}, and from the analyses of the gravitational wave signal from the neutron stars merger GW170817~\cite{Abb17} are shown for comparison. The results shown in Fig.~\ref{fig4} indicate that both the observed maximum mass and corresponding radius of hybrid stars increase considerably with the coupling constant $g$, which is due to that the maximum mass of hybrid stars constrains mostly the EOS of hybrid star matter at densities $2\rho_0 \sim 5\rho_0$~\cite{Liu22}. As shown in Fig.~\ref{fig3}, the hadron-quark mixed phase in most cases occurs at this density region, and thus the properties of SQM affect the maximum mass. As a result, the maximum mass of hybrid stars increases from $1.51$ to $2.06M_{\odot}$ with the increasing $g$. Additionally, we also note that the maximum mass of hybrid stars with parameter sets, $g=5, B^{1/4}=120.0$ MeV and $g=4, B^{1/4}=135.5$ MeV, are very close to the detection result of PSR J0740 + 6620, and the results of hybrid stars in all parameter sets are mostly consistent with the constraints from the pulsars PSR J0030+0451 and the neutron stars merger GW170817.

\begin{figure}[tbh]
\includegraphics[scale=0.27]{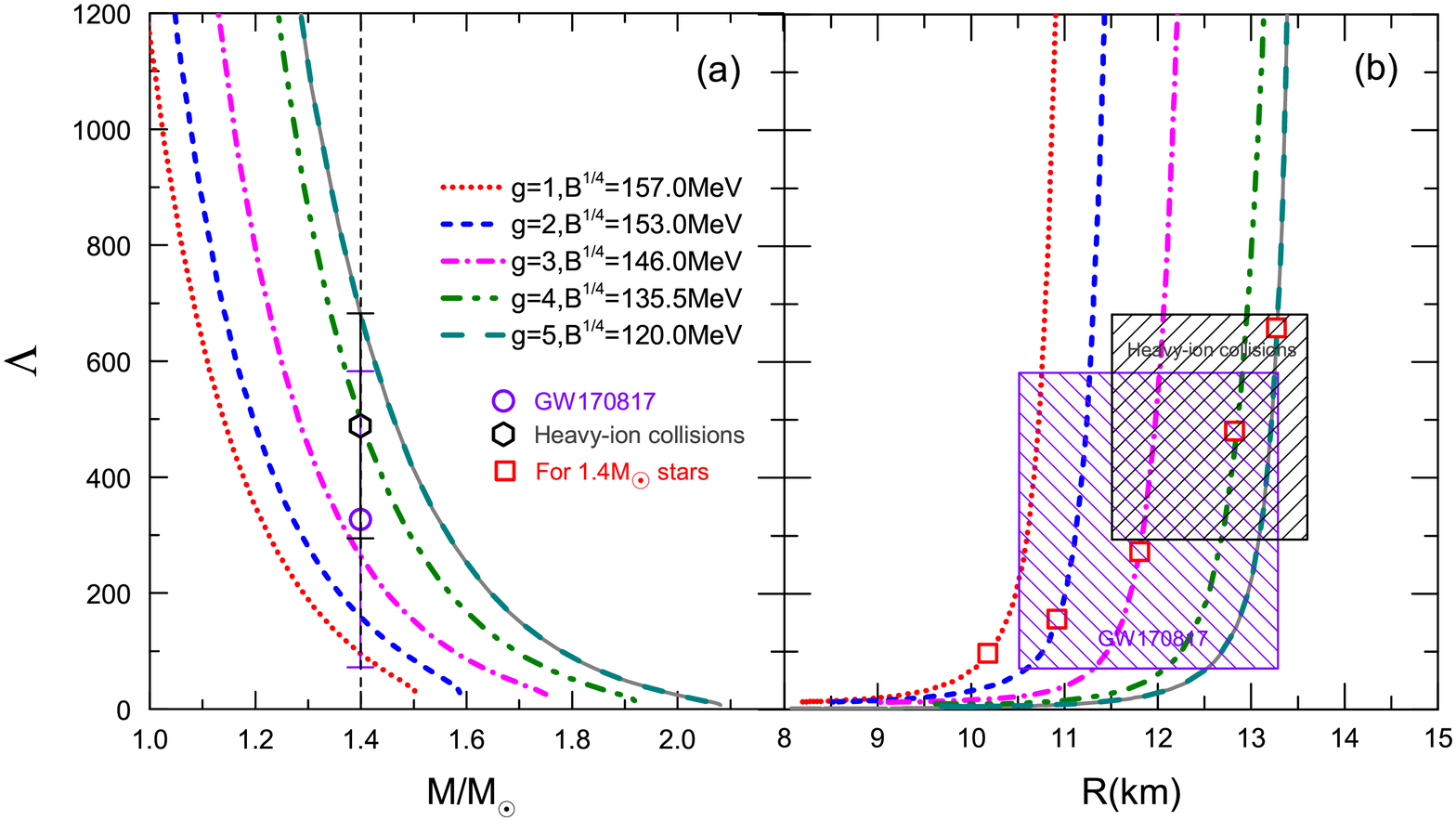}
\caption{(color online) Relations between the dimensionless tidal deformability and the mass as well as between the dimensionless tidal deformability and the radius in hybrid stars with the different coupling constants $g$ for SQM. Both the error bars at $M=1.4 M_{\odot}$ in the left panel and the squared regions in the right panel are derived from the constraints $70 \leq \Lambda_{1.4} \leq 580$ and $10.5 \leq R_{1.4} \leq 13.3$ km based on the improved analysis of GW170817~\cite{Abb18} and the predictions of $292 \leq \Lambda_{1.4} \leq 680$ and $11.5 \leq R_{1.4} \leq 13.6$ km from heavy-ion collisions~\cite{BLi06}. The small red squares indicate the results for hybrid stars with $1.4M_{\odot}$. } \label{fig5}
\end{figure}

After the GW170817 event, much efforts have been devoted to constraining the EOS by comparing various calculations with the range of tidal deformability $70 \leq \Lambda \leq 580$ from the improved analyses of LIGO and Virgo Collaborations. Reams of studies have examined the effects on the nuclear matter~\cite{Kra19,Car19,Zha19}, and some of them have extracted constraints on the slope parameter $L(\rho_0)$, i.e. $L(\rho_0) = 57.7 \pm 19$ MeV~\cite{BLi21}. The measurements of the tidal deformability of neutron stars constrain not only the EOS of dense nuclear matter but also the fundamental strong interactions of quark matter. Shown in Fig.~\ref{fig5} is the dimensionless tidal deformability as functions of mass and radius calculated using the different coupling constants from the quasiparticle model. We see that $\Lambda$ decreases/increases rapidly as the mass/radius of the neutron star increases, which is due to the factor that given the smaller range of allowed radii for massive stars, the spread in the tidal deformability is also naturally much tighter. As expected, increasing  $g$ can stiffen the EOS of hybrid star matter and thus increase the value of the tidal deformability. Both the error bars at $M=1.4 M_{\odot}$ in the left panel and the squared regions in the right panel are derived from the constraints $70 \leq \Lambda_{1.4} \leq 580$ and $10.5 \leq R_{1.4} \leq 13.3$ km based on the improved analyses of GW170817~\cite{Abb18} and the predictions of $292 \leq \Lambda_{1.4} \leq 680$ and $11.5 \leq R_{1.4} \leq 13.6$ km from heavy-ion collisions~\cite{BLi06}. The constraint is derived from the analyses of Ref. [19] under several assumptions, in particular the same EOS for both stars, which is probably not the case if one of them is hybrid stars. Therefore, this constraint is not robust for hybrid stars. In this work, we only consider the range of tidal deformabilities $70 < \Lambda < 580$ as a comparison for hybrid stars. Except for the case $g=1, B^{1/4}=157.0$ MeV, it can be seen that for the canonical mass $\Lambda_{1.4}$ and $R_{1.4} $ with various parameters meet the constraints listed above. In particular, the results of the case, $g=4, B^{1/4}=135.5$ MeV, are approaching the overlapping part of the two constraints.

\begin{figure}[tbh]
\includegraphics[scale=0.26]{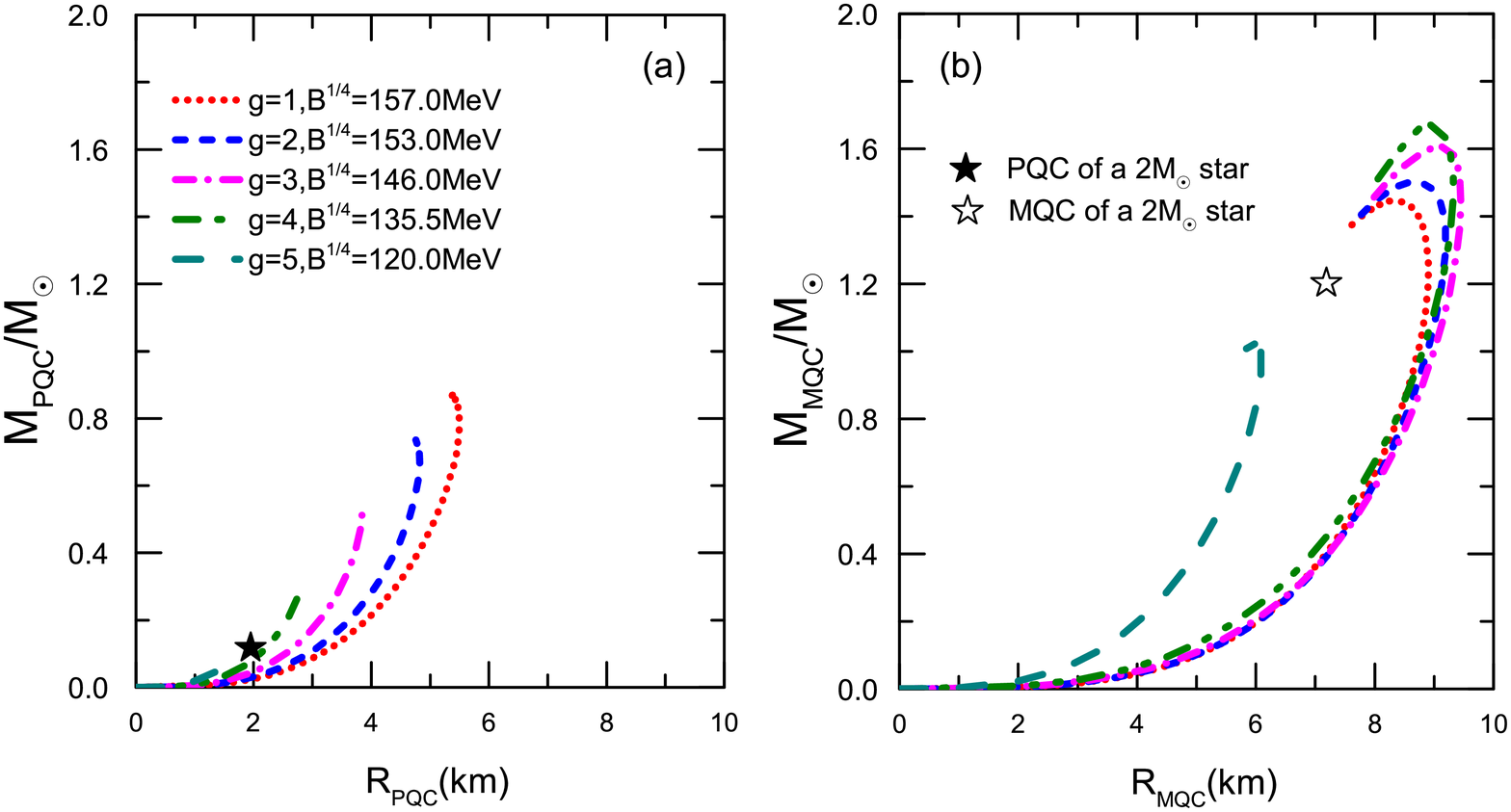}
\caption{(color online) Mass-radius relations of  the pure quark cores (PQC) and the mixed quark cores (MQC) inside hybrid stars based on the quasiparticle model for SQM with different coupling constants $g$. The solid and void stars represent PQC and MQC of a $2M_{\odot}$ hybrid star, respectively. } \label{fig6}
\end{figure}

The hybrid star matter EOS consists of charge-neutral matter in $\beta$-equilibrium that has a hadron-quark phase transition from hadronic to quark matter. To better understand the effects of SQM on the hybrid stars, the relation between the quark-matter core mass $M_{QC}$ and radius $R_{QC}$ is shown in Fig.~\ref{fig6}. Due to the existence of quark matter in the hadron-quark mixed phase, the discussion of the quark-matter core can be divided into two scenarios: the pure quark core (PQC) and the mixed quark core (MQC), where MQC includes the quark phase and mixed phase. The quark matter cores can be determined using the TOV equations by integrating them from the central baryon density to three distinct points: (1) the pressure of onset of pure quark phase where $r=R_{PQC}$ and $M(r)=M_{PQC}$, (2) the pressure of onset of mixed phase where $r=R_{MQC}$ and $M(r)=M_{MQC}$, and (3) zero pressure where $r=R$ and $M(r)=M$. It is clearly seen in Fig.~\ref{fig6} (a) that the coupling constant $g$ has a competitive effect: while the constant $g$ increases the hybrid star mass up to $2M_{\odot}$, it also decreases the PQC mass $M_{PQC}$ and radius $R_{PQC}$. This can be understandable since the onset of the phase transition and the pure quark phase appear at higher densities with increasing $g$ under the Gibbs construction. Particularly, if $g$ is large enough, the onset density of quark matter will be larger than the central density of the massive stars, and thus no PQC can appear in the hybrid stars. With different coupling constants, the maximum mass and radius of PQCs, in current work, are about $0.88M_{\odot}$ and 5.5 km, respectively. However, the mass-radius relation of MQCs in Fig.~\ref{fig6} (b) shows a more complex dependence: the maximum mass and radius of MQCs begin to increase gradually and decrease rapidly after reaching the maximum values with the increase of the constants $g$. The reason for the special dependency is that the mass and radius of MQC with small value $g$ are in close proximity to those of the whole star, which limits the increase of the mass and radius of MQC. For the above $2M_{\odot}$ hybrid stars, the maximum radius of PQCs may only be about $R_{PQC}\approx2.0$ km, while the maximum radius of MQCs (the core where quark matter may appear) can be up to 7.2 km, which is over the quark core radius $R_{QC}=6.5$ km observed in Ref.~\cite{Ann20}.

\begin{figure}[tbh]
\includegraphics[scale=0.31]{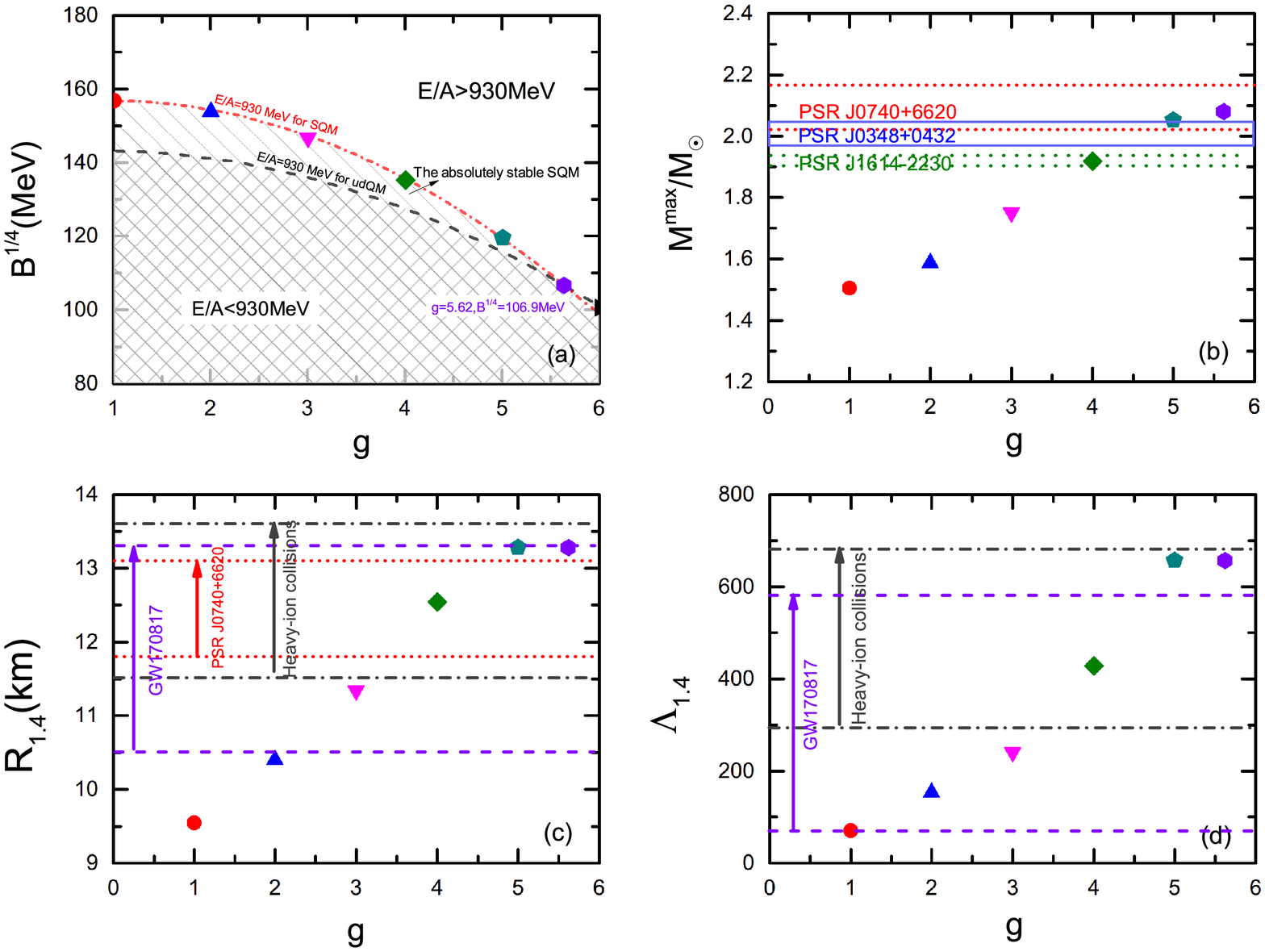}
\caption{(color online) Bag constant $B$ (in panel (a)), maximum mass of hybrid stars (in panel (b)), as well as radius $R_{1.4}$ (in panel (c)) and tidal deformability $\Lambda_{1.4}$ (in panel (d)) for $1.4M_{\odot}$ hybrid stars as functions of the coupling constant $g$ in the quasiparticle model. The horizontal bars in panel (b) indicate the observational constraints of PSR J1614-2230~\cite{Dem10,Arz18}, PSR J0348+0432~\cite{Ant13}, and PSR J0740+6620~\cite{Cro20,Mil21}, and the horizontal bars in panel (c) and (d) are derived from the improved analyses of GW170817~\cite{Abb18} and the predictions from heavy-ion collisions~\cite{BLi06}.} \label{fig7}
\end{figure}

The discussion above has established that changes in $B$ and $g$ can significantly impact the properties of SQM and hybrid stars. Fig.~\ref{fig7} (a) displays the energy per nucleon ($E/A$) of SQM and $ud$QM on the $g-B^{1/4}$ plane. The red dash-dotted line represents an $E/A$ of 930 MeV for SQM, while the black dash line corresponds to an $E/A$ of 930 MeV for $ud$QM. The slash shaded area between the two $E/A=930$ MeV lines represents the region where absolutely stable SQM exists, which decreases as $g$ increases and eventually narrows down to a "point" with $g=5.62$. Hybrid stars containing SQM should break the absolutely stable condition, and the $E/A$ of both $ud$QM and SQM must exceed the lowest $E/A$ value of 930 MeV. As a result, the slash and grid shaded regions represent areas excluded for hybrid stars. Consequently, the colored-marked points on the red dash-dotted line with $g<5.62$, and those on the black dash line with $g>5.62$ signify the minimum values of the bag constants $B$ with different $g$ . Our calculations also reveal that the maximum mass of hybrid stars decreases with increasing $B$ by fixing the constant $g$ since the bag constant $B$ can soften the EOS of SQM through the pressure expression in Eq. (5), while the results of the radius $R_{1.4}$ and tidal deformability $\Lambda_{1.4}$ demonstrate oppositely. Therefore, we can provide the maximum mass, minimum radius $R_{1.4}$ and minimum tidal deformation $\Lambda_{1.4}$ of the hybrid stars with different $g$, as shown in Fig.~\ref{fig7} (b)$\sim$ (d). We can also see from the figures that the maximum mass, radius ($R_{1.4}$), and tidal deformability ($\Lambda_{1.4}$) increase with an increase in $g$. However, it should be noted that different bag constants $B$ when $g>5$ had no impact on the radii and tidal deformabilities of $1.4 M_{\odot}$ hybrid stars since there is no quark matter present inside the stars. For comparison, we depict some constraints in panel (b) from the mass measurements of PSR J1614-2230 ($1.908\pm0.016M_{\odot}$)~\cite{Dem10,Arz18}, PSR J0348+0432 ($2.01\pm0.04M_{\odot}$)~\cite{Ant13}, and PSR J0740+6620 ($2.08\pm0.07M_{\odot}$)~\cite{Cro20,Mil21}, and the horizontal bars in panel (c) and (d) derived from the improved analyses of GW170817~\cite{Abb18} and the predictions from heavy-ion collisions~\cite{BLi06}. It is clearly seen that the cases $g=4$ with various $B$ largely satisfies all the constraints listed.

\begin{figure}[tbh]
\includegraphics[scale=0.24]{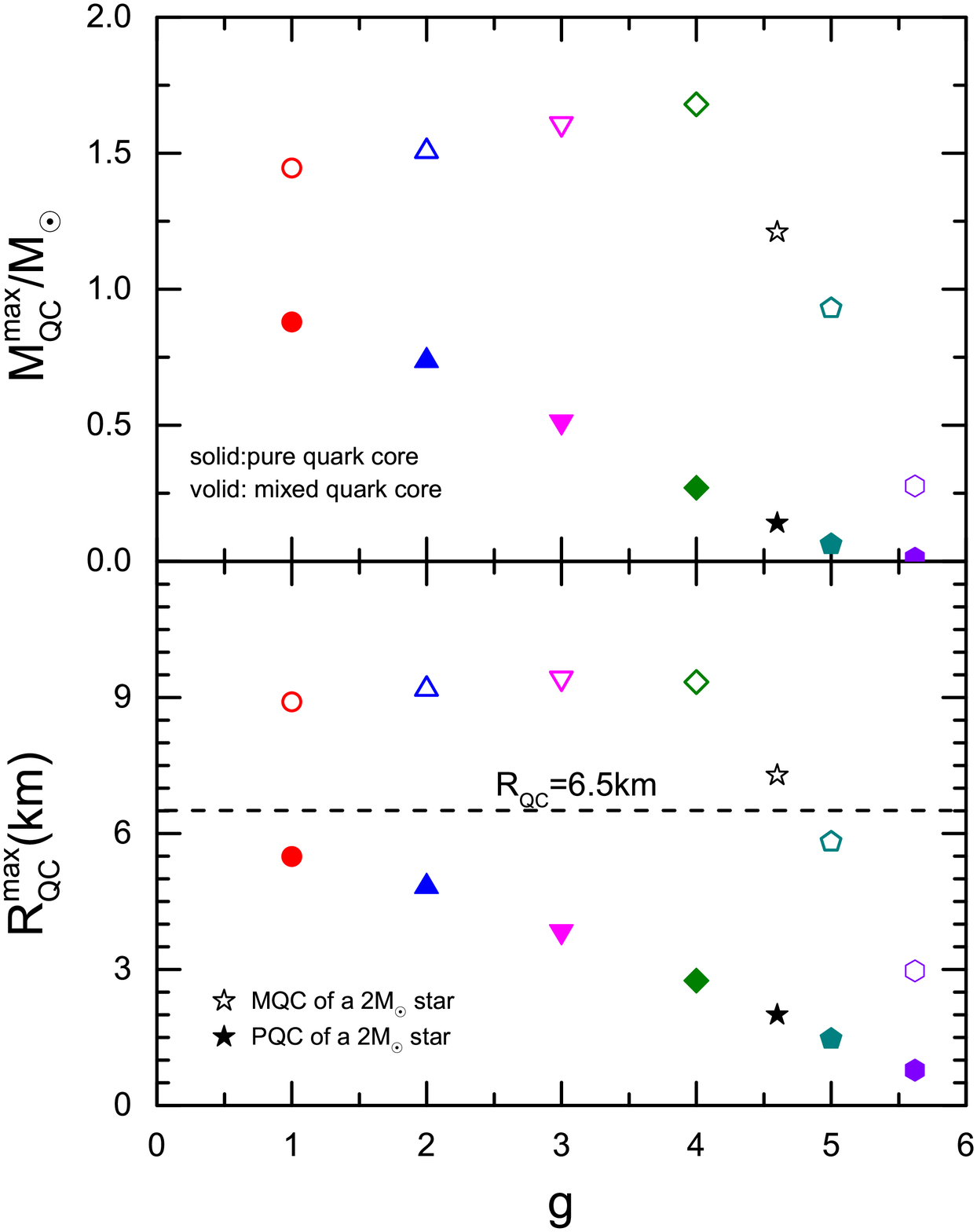}
\caption{(color online) Maximum mass and radius of quark-matter cores inside hybrid stars as functions of the coupling constant $g$ in the quasiparticle model. The solid and void geometries represent the pure quark cores and  mixed quark cores of the hybrid stars respectively, and the dash line represents a 6.5 km quark core observed in Ref.~\cite{Ann20}. } \label{fig8}
\end{figure}

Finally, we show in Fig.~\ref{fig8} the maximum mass and radius of quark-matter cores inside hybrid stars as functions of the coupling constant $g$ with the minimum $B$ in the quasiparticle model.  We can see in the figures that the maximum mass and radius of PQCs decreases with the increment of $g$,  whereas those of MQCs begin to increase gradually and decrease rapidly after reaching the maximum values with the increase of the constants $g$. Different from the complete stars, both maximum mass and radius of PQCs and MQCs have a negative dependence on the bag constant $B$ by fixing the constant $g$, and thus the mass and radius of PQCs and MQCs in Fig.~\ref{fig8} correspond to the maximum values with different $g$.  For the $2M_{\odot}$ stars with the hadron-quark phase transition, the radii of quark-matter cores observed in Ref.~\cite{Ann20} can reach about 6.5 km as shown by the dash line in Fig.~\ref{fig8}. It is clearly seen in Fig.~\ref{fig8} that the PQCs of the massive stars gradually scale down or even nearly disappear. Especially for the extreme value $g=5.62$, the star mass can be up to $2.08M_{\odot}$, whereas the PQC mass almost approaches zero and its radius is also less than 1 km. By contrast, considering the mixed phase, the sizable quark-matter cores ($R_{MQC}>6.5$ km) can appear in $2M_{\odot}$ massive stars.

\section{Summary and Outlook}
\label{viscosity}
 In this work, we have investigated the properties of hybrid stars with the hadron-quark phase transition based on the quasiparticle model. In conclusion, we find that the hybrid star matter EOS is sensitive to the strength of the constants $g$. With increasing the coupling constant $g$ the EOS of hybrid star matter becomes stiffer. Meanwhile, we also note that a step change of the sound velocity occurs in the hadron-quark phase transition, and it is restored with the decrease of nucleon and lepton degrees of freedom in the high density quark phase. Our results agree with the approximate rule following Ref.~\cite{Ann20} that the polytropic index $\gamma \leq 1.75$ can be used as a criterion for separating hadronic from quark matter.  Using the hybrid star matter EOS, we predict the mass-radius relation and tidal deformabilities of hybrid stars as well as the radius and mass information of quark-matter cores. Although the constant $g$ increases the hybrid star maximum mass up to $2.08M_{\odot}$, it also decreases the mass and radii of quark-matter cores. However, considering the quark matter in the mixed phase, we confirm that the sizable quark-matter core $R_{QC}=7.2$ km can appear in $2M_{\odot}$ massive stars.

 The hypothesis of absolutely stable SQM (or "Witten hypothesis") suggests that a hybrid star containing a sufficient amount of SQM in its core will rapidly convert into a strange quark star. The SQM in hybrid stars therefore should break the absolutely stable condition, and the energy per nucleon ($E/A$) of both $ud$QM and SQM must exceed the lowest energy per nucleon 930 MeV.  In the work, we present the energy per nucleon ($E/A$) of SQM and $ud$QM and regions excluded for hybrid stars on the $g-B^{1/4}$ plane. We find that the maximum mass of hybrid stars decreases with increasing $B$ since the bag constant $B$ can soften the EOS of SQM, while the results of the radius $R_{1.4}$  and tidal deformability $\Lambda_{1.4}$ for a $1.4M_{\odot}$ star demonstrate oppositely. Different from the complete stars, both maximum mass and radius of quark-matter cores have a negative dependence on the bag constant $B$.  As a result, we provide the maximum mass, minimum radius $R_{1.4}$  and tidal deformation $\Lambda_{1.4}$  of the hybrid stars as well as the maximum mass and radius of the quark matter core with different $g$ values within the allowable regions ($E/A>930$ MeV) on the $g-B^{1/4}$ plane. For comparison, we also display some constraints from astrophysical observations and heavy-ion experiments, which comprise the multi-messenger data set for our analyses on the properties of hybrid star matter. In addition, some of the new discoveries and observations provide more rigorous constraints on SQM, or may also contain some new physics. For an example, the newly discovered compact binary merger GW190814 with a secondary component of mass $2.50\sim2.67M_{\odot}$, which can be reproduced by a superfast pulsar~\cite{Zha20} or quark star~\cite{Zha21,Chu21}. Those observations of massive stars can also be used to understand the properties of the QCD phase transition. To further explore the QCD phase structure and search for the signal of the QCD critical point, experimental programs such as the beam-energy scan (BES) at RHIC were proposed. The promising results are available to provide more constraints on the EOSs of SQM, and are helpful in the understanding of the QCD phase structure.

\begin{acknowledgments}
This work is supported by the National Natural Science Foundation of China under Grants No. 12205158 and No. 11975132, and the Shandong Provincial Natural Science Foundation, China Grants No. ZR2021QA037, No. ZR2022JQ04, No. ZR2019YQ01, and No. ZR2021MA037.
\end{acknowledgments}

\end{document}